\documentclass[aps, prd, amsmath, floats, floatfix, twocolumn, nofootinbib]{revtex4-2}
\usepackage{graphicx}
\usepackage{color}
\usepackage{latexsym}
\usepackage{amsthm,amsmath,amssymb}
\usepackage{mathrsfs}
\usepackage{amssymb}
\usepackage{booktabs}
\usepackage{makecell}

\usepackage{tikz}  
\usetikzlibrary{arrows,shapes}
\usetikzlibrary{trees}
\usetikzlibrary{matrix,arrows} 
\usetikzlibrary{positioning}				
\usetikzlibrary{calc,through}				
\usetikzlibrary{decorations.pathreplacing}  
\usepackage{pgffor}							
\usetikzlibrary{decorations.pathmorphing}	
\usetikzlibrary{decorations.markings}

\newcommand{\be}{\begin{eqnarray}}
\newcommand{\ee}{\end{eqnarray}}

\newcommand{\bea}{\begin{eqnarray}}
\newcommand{\eea}{\end{eqnarray}}

\begin{document}

\title{On the analytic extension 
of regular rotating black holes}

\author{Tian Zhou}
\email{zhout6@mail.sustech.edu.cn}

\author{Leonardo Modesto}
\email{lmodesto@sustech.edu.cn}

\affiliation{Department of Physics, Southern University of Science
and Technology, Shenzhen 518055, China}

\begin{abstract}

We hereby focus on the analytic geodesic extension of several regular rotating black holes (RRBHs) obtained through out the Newman-Janis algorithm starting from some popular spherically symmetric regular black holes. 
It turns out that if the metric is not an even function of Boyer-Lindquist radial coordinate $r$, similarly to the Kerr spacetime, the metric has to be extended to negative values of $r$ to ensure the analyticity of the geodesic equations (and in turn of the geodesics). Therefore, some of the extended RRBHs considered in this paper, such as the rotating Hayward black hole, are geodetically incomplete because they are singular somewhere for $r<0$, and non-analytic at the ring located in $(r=0, \theta=\pi/2)$. Conversely, other spacetimes, like nonlocal black holes, can be analytically extended to negative $r$. 

However, the real issue shows up at the ring, where, 
unfortunately, all the RRBHs studied in this paper fail to be regular. Indeed, at the ring, the Kretschmann invariant is finite but nonanalytic, while the higher derivative curvature invariants are divergent.

In order to avoid such catastrophe, we propose a modification of the RRBHs in which the angular momentum is promoted to a function of the radial coordinate. According to our proposal, the angular momentum vanishes for $r\rightarrow 0$ and the ring shrinks to a point. Therefore, the regularity properties of the regular spherically symmetric black holes are recovered for $r \rightarrow 0$.

%
%

\end{abstract}

\maketitle

\section{Introduction}
A general spherically symmetric spacetime line element 
is given by the following line element, 
\be
ds^2 =  f(r) dt^2 - g^{-1}(r) dr^2 - h(r) d \Omega^{(2)}\, . 
\label{SchLINE}
\ee
For a black hole in Schwarzschild coordinates, the function $f(r)$ and $g(r)$ are equal, while $h(r)=r^2$. Hence, the metric can be expressed in terms of one function $m(r)$, namely\footnote{We here work in Planck units, namely the Newtonian gravitational constant, the speed of the light, and the Plank's constant are: $c= \hbar = G =1$.}
\be\label{Schwarzlike}
f(r) = 1-\frac{2m(r)}{r}\, , 
\ee
where the function $m(r)$ is a function of the radial coordinate $r$. For the Schwarzschild black hole, $m(r)$ is constant, and we get a singularity in $r=0$. In order to cure the singularity of the Schwarzschild black hole, many regular spherically symmetric spacetimes have been proposed \cite{Bardeen, Hayward:2005gi, Dymnikova:1992ux, Bonanno:2000ep, Modesto:2009ve, Nicolini:2005vd, Modesto:2010uh, Burzilla:2020utr, Modesto:2008im, Culetu:2013fsa, Culetu:2014lca, Xiang:2013sza, Simpson:2019mud}, even though the underlying dynamics for some of these regular black holes are still unclear. In particular, for the Hayward black hole \cite{Hayward:2005gi}, which is a popular model for spherically symmetric regular black hole, the function $m(r)$ reads:
\be\label{hayward}
m(r)= \frac{Mr^3}{r^3+L^3}\, ,
\ee
where $M$ is a constant of dimension mass dimension and $L$ is a constant of length dimension. 
For the Hayward metric, $1 - f(r)\propto r^2$ in the limit $r\rightarrow 0$. Hence, the metric has a de-Sitter core at the center of the black hole, and one can check that the curvature invariants, such as the Kretschmann scalar $R_{\mu\nu\rho\sigma}R^{\mu\nu\rho\sigma}$, are regular at $r=0$ because 
$m(r)$ is at least cubic, i.e. $r^3$, when $r\rightarrow0$.

Therefore, the singularity problem seems to disappear in spacetimes having the same behavior of the Hayward black hole near $r\sim 0$. 

However, in Ref.\cite{Zhou:2022yio} has been shown that the analytic extension of the Hayward spacetime is actually singular for $r <0$ in $r=-L$. Moreover, the antipodal extension is not doable because the geodesic equations are not analytic.
Furthermore, in \cite{Giacchini:2021pmr} it was proved that higher-derivative curvature invariants diverge
for some regular black holes, 
 thus these metrics might be filtered out by the finite action principle if the gravitational action contains higher-derivative curvature invariants. The two results of the above provide constraints on the form of the metric for a spherically symmetric regular black hole. 

In order to generalize the results in \cite{Zhou:2022yio}, in this paper, we focus on several supposed to be regular rotating black holes, which are physically relevant because of the universally recognized existence of rotating black holes in the Universe. 
The regular rotating black holes (RRBH) can be obtained from the spherically symmetric ones by making use of the Newman-Janis algorithm (for more details see \cite{newman, Drake:1998gf}). 


For instance, by applying the Newman-Janis transformation to the Schwarzschild metric, one can obtain the Kerr metric in Boyer-Lindquist coordinates
\begin{align}\label{kerr}
ds^2& = \left(1-\frac{2Mr}{\Sigma} \right)dt^2 + \frac{4aMr\rm{sin}^2\theta}{\Sigma}dt d\phi - \frac{\Sigma}{\Delta}dr^2 \nonumber\\
&- \Sigma d\theta^2 - {\rm{sin}}^2\theta \left(r^2+a^2+\frac{2a^2Mr{\rm{sin}}^2\theta}{\Sigma} \right)d\phi^2 ,
\end{align}
where $\Sigma=r^2+a^2{\rm cos}^2\theta$ and $\Delta = r^2-2Mr+a^2$. The parameter $a$ is a constant for the Kerr spacetime, and it can be interpreted as the angular momentum of the rotating black hole, namely $a=J/M$. When $a=0$, the Kerr metric reduces to the Schwarzschild one, while a singularity is located at $(r=0,\theta=\pi/2)$, which is known as the ring singularity. 

Regular black holes have been proposed starting from spherically symmetric regular black holes and applying the Newman-Janis algorithm \cite{Bambi:2013ufa, Modesto:2010rv, Toshmatov:2014nya,Ghosh:2014pba}. 
As a popular example, the rotating Hayward spacetime has the same form of (\ref{kerr}), except for replacing the constant mass $M$ 
with the mass function $\tilde{m}(r)$ (see next section). 

It is common that the rotating Hayward black hole is regular everywhere because the ring singularity is inflated to a {\em de-Sitter belt.} However, in this letter, we point out that the rotating Hayward black hole should be analytically continued to negative values of coordinate $r$, like for the Kerr spacetime, and then the maximally extended black hole will be singular again. The result is based on the analyticity of the maximal extension for geodesics, which is usually required in general relativity because it ensures the uniqueness of the extension in order to preserve predictability. 
Hence, it is necessary to investigate the analytic properties of the supposed to be regular spacetimes and revisit the singularity issue for the RRBHs. The conclusions will also be a constraint to select the correct spherically symmetric regular black holes as the proper one able to induce a regular rotating black hole.

\section{Newman-Janis algorithm}
We here briefly review the Newman-Janis algorithm applied to the metric (\ref{SchLINE}) (see \cite{Bambi:2013ufa} for more details). 
The first step consists in introducing the advanced null coordinate $u$ defined by 
\be
du=dt - dr/f(r)\, .
\ee
The second and crucial step consists in promoting the real coordinates $(r,u)$ 
 to complex coordinates, namely 
\be
r \,\, \rightarrow \,\, r' = r+i\, a\, {\rm cos}\theta \, ,\quad u \,\, \rightarrow \,\, u' = r - i\, a\, {\rm cos}\theta\, ,
\label{complex}
\ee
%
The functions $\tilde{f}(r,\theta)$ and $\tilde{h}(r,\theta)$ on the complex domain are obtained replacing $r$ with $r^\prime$, according to
(\ref{complex}), in $f(r)$ and $h(r)$ (see (\ref{SchLINE})) and imposing the metric to be real. 
Hence, applying the following coordinate transformation, 
\be
du = dt' + F(r)dr\, ,\quad d\phi = d\phi' + G(r)dr\, ,
\ee
where the following functions can only depend on the radial coordinate $r$, 
\be\label{FG}
F(r)=\frac{\tilde{h}+a^2{\rm sin}^2\theta}{\tilde{f}\tilde{h}+a^2{\rm sin}^2\theta}\, ,\quad G(r)=\frac{a}{\tilde{f}\tilde{h}+a^2{\rm sin}^2\theta}\, ,
\ee
we obtain the rotating black hole in Boyer-Lindquist coordinates, i.e. 
\begin{align}\label{rbh}
ds^2 = &\tilde{f}dt^2 + a\,{\rm sin}^2\theta (1-\tilde{f})dt d\phi - \frac{\tilde{h}}{\tilde{f}\tilde{h}+a^2{\rm sin}^2\theta}dr^2 \nonumber\\
&- \tilde{h} d\theta^2 - {\rm{sin}}^2\theta [\tilde{h}+a^2{\rm sin}^2\theta (2 - \tilde{f})] d\phi^2 ,
\end{align}
 which has one off-diagonal component $g_{t\phi}$. 
 %

 
The simplest complexification to make the Newman-Janis algorithm to work consist in the following replacement \cite{Bambi:2013ufa}:
\be
&&
\frac{1}{r} \,\, \rightarrow \,\, \frac{1}{2}\left( \frac{1}{r'}+\frac{1}{\bar{r}'} \right)=\frac{r}{\Sigma} \, , \quad r^2 \,\, \rightarrow \,\, r'\bar{r}' = \Sigma\, , \nonumber \\
&&
\tilde{m}(r) = m(r)\, .
\ee
Therefore, the functions $\tilde{f}(r,\theta)$ and $\tilde{h}(r,\theta)$ in the metric (\ref{rbh}) are:
\be\label{fh}
\tilde{f}(r') =1-\frac{2\tilde{m}(r)r}{\Sigma}\, ,\quad \tilde{h}(r,\theta) = \Sigma\, .
\ee
Notice that the chosen complexification (\ref{fh}) allow to recover the Kerr metric (\ref{kerr}) by simply taking $\tilde{m}(r)=M$.


\section{Analytic continuation beyond the disk}

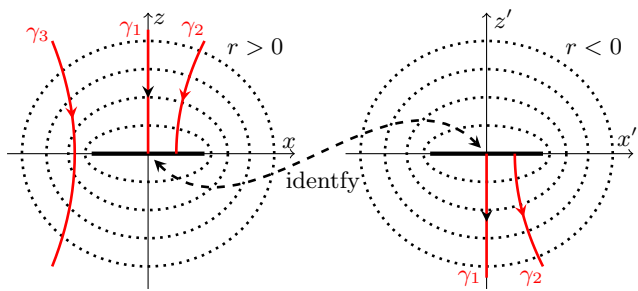
\begin{figure}[hbt]
\centering
\begin{tikzpicture}[line width=1 pt, scale=0.75]

\draw[->,thin] (-2.5,0)--(2.6,0); \node at (2.5,0.2) {$x$};
\draw[->,thin] (0,-2.4)--(0,2.5); \node at (0.2,2.4) {$z$};
\draw[ultra thick] (-1,0)--(1,0);
\draw [dotted] (0,0) ellipse ( 1.118 and 0.5);
\draw [dotted] (0,0) ellipse ( 1.414 and 1);
\draw [dotted] (0,0) ellipse ( 1.803 and 1.5);
\draw [dotted] (0,0) ellipse ( 2.236 and 2);
\draw [draw=red, postaction={decorate}, decoration={markings, mark=at position .55 with {\arrow[draw=red]{stealth}}}] (0,2.2)--(0,0);
\draw [domain=2:0, red, postaction={decorate}, decoration={markings, mark=at position .55 with {\arrow[draw=red]{stealth}}}] plot({(\x)^2/8+0.5}, \x);
\draw [domain=2:-2, red, postaction={decorate}, decoration={markings, mark=at position .35 with {\arrow[draw=red]{stealth}}}] plot({-(\x)^2/10-1.3}, \x);
\draw[stealth-stealth,dashed] (315:0.15) to[out = 315, in=135] (5.9,0.1);
\node at (3.1,-0.5) {identfy}; \node at (1.9,1.9) {$r>0$}; \node at (-0.3,2.2) {\color{red}$\gamma_1$}; \node at (0.8,2.2) {\color{red}$\gamma_2$}; \node at (-1.95,2.1) {\color{red}$\gamma_3$};

\begin{scope}[shift={(6,0)}]
\draw[->,thin] (-2.5,0)--(2.6,0); \node at (2.5,0.3) {$x’$};
\draw[->,thin] (0,-2.4)--(0,2.5); \node at (0.3,2.4) {$z'$};
\draw[ultra thick] (-1,0)--(1,0);
\draw [dotted] (0,0) ellipse ( 1.118 and 0.5);
\draw [dotted] (0,0) ellipse ( 1.414 and 1);
\draw [dotted] (0,0) ellipse ( 1.803 and 1.5);
\draw [dotted] (0,0) ellipse ( 2.236 and 2);
\draw [draw=red, postaction={decorate}, decoration={markings, mark=at position .55 with {\arrow[draw=red]{stealth}}}] (0,0)--(0,-2.2);
\draw [domain=0:-2, red, postaction={decorate}, decoration={markings, mark=at position .55 with {\arrow[draw=red]{stealth}}}] plot({(\x)^2/8+0.5}, \x);
\node at (1.9,1.9) {$r<0$};  \node at (-0.3,-2.2) {\color{red}$\gamma_1$}; \node at (0.8,-2.2) {\color{red}$\gamma_2$};
\end{scope}

\end{tikzpicture}
\caption{Analytic continuation of the geodesics for regular rotating black holes.}
\label{fig1}
\end{figure}

In this section, we show that most of the RRBHs can be extended analytically to negative values of the radial coordinate. 
Let us consider null geodesics that obey the following equation, 
\be\label{guu}
g_{tt}\dot{t}^2 + 2g_{t\phi}\dot{t}\dot{\phi} + g_{rr}\dot{r}^2 + g_{\theta\theta}\dot{\theta}^2 + g_{\phi\phi}\dot{\phi^2} = 0\, ,
\ee 
(which comes from $ds^2 = 0$) 
where the dot denotes the derivative respect to the affine parameter $\lambda$. Since the metric (\ref{rbh},\ref{fh}) is static and axisymmetric, we have the following Killing vectors, 
\be
\xi^\mu=(1,0,0,0)\, , \quad \eta^\mu=(0,0,0,1)\, ,
\ee
to which correspond the following conserved quantities, 
\begin{align}\label{conserved}
e &\equiv \xi^\mu u^\nu g_{\mu\nu}=g_{tt}\dot{t}+g_{t\phi}\dot{\phi} \, , \nonumber\\
l &\equiv \eta^\mu u^\nu g_{\mu\nu}= g_{\phi t}\dot{t} + g_{\phi\phi}\dot{\phi} \, .
\end{align}
The equations (\ref{guu},\ref{conserved}) govern the motion of massless particles. Here, we would like to consider the special case of massless probe particles moving along the rotation axis (see the geodesic $\gamma_1$ in FIG.\ref{fig1}), namely we take $\theta=0$ and $\dot{\theta}=0$. Therefore, the equation of motion (\ref{guu},\ref{conserved}) simplify to:
\begin{align}\label{eomtheta0}
 \left(1-\frac{2\tilde{m}(r)r}{\Sigma} \right)\dot{t}^2 - \frac{\Sigma}{\Delta}\dot{r}^2 = 0\, ,  \nonumber \\
 e=\left(1-\frac{2\tilde{m}(r)r}{\Sigma} \right)\dot{t} \, , \quad l=0 \, ,
\end{align}
where we replaced $g_{t\phi}=g_{\phi\phi}=0$, since we are considering $\theta=0$, and
\be
\frac{\Sigma}{\Delta} = \frac{r^2+a^2}{r^2-2\tilde{m}(r)r+a^2}=\left(1-\frac{2\tilde{m}(r)r}{\Sigma} \right)^{-1}\, .
\ee
Finally, the geodesic equation for massless particles is given by:
\be\label{dott2}
\dot{r}^2 = e^2 = {\rm const.} \, , \quad  \dot{t}^2 (r) = \left( 1-\frac{2\tilde{m}(r)r}{r^2+a^2} \right)^{-2}\, .
\label{rdotte}
\ee
Now we consider the analytic extension of the geodesics after they reach the interior of the disk located in $r=0$. 

For the sake of simplicity, we take ${\rm const.} =1$ in (\ref{rdotte}). 
If the spacetime is only defined in $r\ge 0$, the geodesic should trivially pierce through the disk and then reach the opposite side of the disk. We call such extension: {\emph {antipodal continuation}}. Therefore,  the function $\dot{r}$ should suddenly changes the sign at $r=0$, namely $\dot{r}=-1\rightarrow +1$. Indeed,  $\dot{r}$ is positive for particles moving towards the ring, while it is negative for particles moving out of the ring.
This usually leads to non-analyticity of the geodesic equation $\dot{t}(\lambda)$.

Let us consider the rotating Hayward metric as a typical example to explicitly show the issue. Near $r=0$, the function $\dot{t}^2(r)$ can be expanded as
\be\label{expanddott2}
\dot{t}^2 (r) = 1 + \frac{4M}{a^2L^3}r^4 - \frac{4M}{a^4L^3}r^6 - \frac{4M}{a^2L^6}r^7 + \mathcal{O}(r^7) . 
\ee
Replacing the variable $r$ by $\lambda_0-\lambda$ and $\lambda-\lambda_0$ before and after arriving in $r=0$ respectively (where $\lambda_0$ is the affine parameter when the geodesic reaches the disk $r=0$), the $r^7$ term of the function $\dot{t}^2(\lambda)$ flips the sign for $\lambda=\lambda_0$, and thus the geodesic is not smooth. Hence, to preserve the analyticity of the geodesics, we are forced to extend the above geodesic to negative values of $r$. In this way, the sign of $\dot{r}$ does not change when crossing the disk, and the geodesic equation is analytic.

We can also consider general geodesics crossing the equatorial plane $z=0$ within the ring $(r=0,\theta=\pi/2)$  (see the geodesic $\gamma_2$ in FIG.\ref{fig1}). 
From the conserved quantities in (\ref{conserved}), the geodesics satisfy the following equation, 
\be\label{dott}
\dot{t}=\frac{e g_{\phi\phi}-l g_{t\phi}}{g_{tt}g_{\phi\phi} - g_{t\phi}^2}\, 
\ee
Therefore, taking multiple derivative of $\dot{t}$, we get:
\be
\hspace{-0.2cm} 
\left( \frac{d}{d\lambda} \right)^{\! n}  \!\! \dot{t} =\left[ \dot{r}^n \left( \frac{\partial}{\partial r} \right)^{\! n} + \dot{\theta}^{n} \left( \frac{\partial}{\partial \theta} \right)^{\! n} \right] \frac{e g_{\phi\phi}-l g_{t\phi}}{g_{tt}g_{\phi\phi} - g_{t\phi}^2} .
\ee
Similarly to the previous case, if we do not extend the spacetime to $r<0$, $\dot{r}$ changes sign when the particles pass through the disk. 
The sudden change of the sign of $\dot{r}$ makes $(d/d\lambda)^n\dot{t}$, with odd power $n$, discontinuous. Therefore, the geodesic equation $\dot{t}$ is non-analytic. 
On the other hand, if the formula (\ref{dott}) for $\dot{t}$ is an even function of the coordinate $r$, then all the terms $\partial_r^n \dot{t}$ with odd power $n$ vanish and $(d/d\lambda)^n\dot{t}$ is continuous. 

In order to ensure $\dot{t}$ to be an even function of $r$ for all values of $e$ and $l$, the metric components $g_{\phi\phi}$, $g_{t\phi}$, and $g_{tt}$ should be even functions of $r$. 
For rotating black holes described by the metric (\ref{rbh},\ref{fh}), the above conditions impose the function $\tilde{f}(r,\theta)$ to be an even function of $r$, or equivalently, the mass function $\tilde{m}$ to be an odd function of $r$. 

Clearly, the mass function (\ref{hayward}) of the rotating Hayward metric is not an even function of $r$, thus the antipodal continuation is not analytic, and the geodesics going towards the ring must be continued for $r<0$, as schematically shown in FIG.\ref{fig1}. 

In general, for the rotating black holes defined by the metric $(\ref{rbh},\ref{fh})$, we can make the following statement: \emph {If the mass function $\tilde{m}(r)$ is an analytic but not an odd function of $r$, the geodesics reaching the disk must be analytically extended to the region with negative values of $r$.}

It turns out that for the popular Kerr-like metrics shown in the TABLE. \ref{massfunction}, the functions $\tilde{m}(r)$ is not an odd function of $r$, and the spacetime must be extended to $r<0$ in order to preserve the analyticity of geodesics. 


\begin{table}[hbt]
\centering
\caption{Some mass function examples}
\label{massfunction}
\begin{tabular}{cc}
   \toprule
   metric & $\tilde{m}(r)$  \\
   \midrule
   Kerr & $\tilde{m}(r)=M$ \\
   Hayward \cite{Hayward:2005gi} & $\tilde{m}(r)=M\frac{r^3}{r^3+L^3} $  \\
   Bardeen \cite{Bardeen} & $ \tilde{m}(r)=M\left(\frac{r^2}{r^2+L^2}\right)^{3/2} $ \\
   Dymnikova \cite{Dymnikova:1992ux}  & $\tilde{m}(r)=M (1-{\rm e}^{-r^3/L^3})$ \\
   \makecell{Noncommutative \cite{Nicolini:2005vd,Modesto:2010rv}\\ and nonlocal \cite{Modesto:2010uh, Buoninfante:2022ild} } & $\tilde{m}(r)=M\left(1-\frac{\Gamma(3/2,r^2/4\alpha)}{\Gamma(3/2)}\right)$ \\
   RG improved \cite{Bonanno:2000ep}  & $\tilde{m}(r)=M\frac{r^3}{r^3+\alpha r +\beta}$ \\
   \bottomrule
\end{tabular}
\end{table}

\section{Geodesic incompleteness inside the ring}

As we have shown in the previous section, many regular rotating black holes must be analytically extended to $r<0$. Therefore, we have to revisit the regularity of such extended spacetimes. Taking again the rotating Hayward metric as an example, it is easy to see that the rotating metric with mass function (\ref{hayward}) is singular in $r=-L$. One can verify that the Kretschmann curvature invariant of the 
Hayward metric in the limit $r\rightarrow -L$ is singular, namely
\be
&&\hspace{-0.4cm}
\lim_{r \rightarrow -L}R_{\mu\nu\rho\sigma}R^{\mu\nu\rho\sigma}= 
\lim_{r \rightarrow -L}\frac{16M^2L^4}{9(L^2+a^2)^2(r+L)^6} = +\infty \, , \nonumber \\
&&
\ee
For the null geodesic described by the equation (\ref{eomtheta0}), we have $\dot{r}= {\rm const}$. The amount of affine parameter to reach $r=-L$ is therefore finite. Therefore, the maximally analytic continuation of the rotating Hayward spacetime is geodetically incomplete.

Similarly, one can check the geodesic (in-)completeness of other RRBHs in Table.\ref{massfunction}. The mass function of the renormalization group(RG) improved black hole\cite{Bonanno:2000ep}, which depends on two positive parameters $\alpha$ and $\beta$, is also singular in $r<0$, like the Hayward metric. The rotating black hole with RG improved mass function is therefore geodesically incomplete. 

For the mass functions of the Bardeen\cite{Bardeen}, and black holes obtained from noncommutative geometry\cite{Modesto:2010rv}, the corresponding metrics are regular everywhere in $(-\infty, +\infty)$.

In non-local gravity, there are no exact black hole solutions so far, but some approximate solutions can be obtained for theories defined in terms of some special form factors. A papular example is given by the Gaussian form factor \cite{Modesto:2010uh}, for which the metric has the same form as the black hole metric in noncommutative spacetime geometry. For this kind of nonlocal black hole, there is no singularity in the region $r<0$.

For the rotating Dymnikova black hole \cite{Dymnikova:1992ux}, the metric is divergent for $r\rightarrow - \infty$. One can also check that the Kretschmann invariant is singular in the $r\rightarrow -\infty$ limit. Hence, we have to investigate the geodesic completeness of the spacetime. From the geodesic equation $\dot{r}= {\rm const.}$ for null geodesics with $\theta=0$, the amount of affine parameter for massless particles to reach $r=-\infty$ is infinite. However, we will see that massive probe particles will reach $r=-\infty$ in finite proper time. Indeed, let us consider massive particles for $\theta=0$ and $\dot{\theta}=0$. The proper time parametrization implies:
\be
\left(1-\frac{2\tilde{m}r}{\Sigma} \right)\dot{t}^2 - \frac{\Sigma}{\Delta}\dot{r}^2 = 1 \, .
\ee
Using the conserved quantity $e=g_{tt}\dot{t}$, we can express the above solution equation as:
\be
\dot{r}^2=e^2-1+\frac{2\tilde{m}r}{\Sigma}\, .
\ee
In the $r\rightarrow -\infty$ limit, we have $\tilde{m}\rightarrow -M{\rm e}^{-r^3/L^3}$ and 
\be
\dot{r}^2\rightarrow -\frac{M{\rm e}^{-r^3/L^3}}{r}\, .
\ee
Therefore, the amount of proper time to reach $r=-\infty$ is:
\be
\Delta\tau=\int_{-\infty}^{r_0} \frac{1}{\sqrt{\dot{r}^2}} \sim \int_{-\infty}^{r_0} \ \sqrt{-r} \ {\rm e}^{r^3/2L^3}\, ,
\ee
which is finite. Thus, the rotating Dymnikova black hole is incomplete for massive particles.


\section{Geodesics incompleteness at the ring}

The analytic extension of the geodesics passing through the interior of the ring located at $(r=0, \theta=\pi/2)$ has been studied in the previous section and illustrated in FIG.\ref{fig1}. 
In this section, we consider the extension of the geodesics reaching exactly the ring, namely the boundary of the disc. For the sake of simplicity, we focus on null geodesics moving on the equatorial plane $z=0$ and approaching the ring from outside, namely $\theta=\pi/2$ and $\dot{\theta}=0$. Again, since the conserved quantities (\ref{conserved}), we have:
\be\label{dottdotphi}
\dot{t}=\frac{e g_{\phi\phi}-l g_{t\phi}}{g_{tt}g_{\phi\phi} - g_{t\phi}^2}\, ,\quad \dot{\phi}=\frac{l g_{tt}-e g_{t\phi}}{g_{tt}g_{\phi\phi} - g_{t\phi}^2}\, .
\ee
Plugging (\ref{dottdotphi}) into the equation $ds^2 = 0$, namely 
\be
g_{tt}\dot{t}^2 + 2g_{t\phi}\dot{t}\dot{\phi} + g_{rr}\dot{r}^2 + g_{\phi\phi}\dot{\phi^2} = 0\, ,
\ee
finally, we get: 
\be\label{eomdotr}
\dot{r}^2 = \frac{ e^2g_{\phi\phi} - 2el g_{t\phi} + l^2 g_{tt} }{ g_{rr}(g_{t\phi}^2 - g_{tt}g_{\phi\phi})}\, .
\ee
Assuming $\tilde{m}\sim r^n$ with $n\ge 3$ for $r$ close to zero, the asymptotic behavior of the metric components in the limit $r\rightarrow0$ reads:
\begin{align}
&g_{tt}=1-\frac{2\tilde{m}}{r}\rightarrow 1 \, , \quad \ g_{t\phi}=\frac{2a\tilde{m}}{r} \,\, \rightarrow \,\, 2ar^{n-1}\, , \nonumber\\
&g_{rr}=-\frac{r^2}{r^2-2\tilde{m}r+a^2} \,\, \rightarrow \,\, -\frac{r^2}{a^2} \, ,\nonumber\\
&g_{\phi\phi}=-\left( r^2 +a^2 +\frac{2a^2\tilde{m}}{r} \right) \,\, \rightarrow \,\, - a^2 \, .
\end{align}
Therefore, the function $\dot{r}^2$ has the following limit, 
\be
\dot{r}^2 \,\, \xrightarrow{r\rightarrow 0}  \,\, \frac{l^2}{r^2} \, .
\ee
Therefore, $\dot{r}$ diverges when the massless particle approaches the ring, and the affine parameter to reach the ring, namely $\Delta\lambda=\int_{r_0}^0 dr/\dot{r}$, turns out to be finite. Hence, we are forced to extend the geodesics beyond the ring. 
It seems that one can not do it because $\dot{r}$ is singular in $r=0$. However, using the polar coordinates $(\rho,\phi)$, in which $\rho=\sqrt{r^2+a^2}=\sqrt{x^2+y^2}$, we find that the limit of $\dot{\rho}$ is actually finite, namely
\be
\dot{\rho}=\dot{r}\frac{d\rho}{dr}=\dot{r}\frac{r}{\rho} \,\, \xrightarrow{r\rightarrow0}  \,\, -\frac{l}{a} \, .
\ee
Meanwhile, $\dot{\phi}$ is also finite in the limit $r\rightarrow0$, i.e.
\be\label{eomdotphi}
\dot{\phi} \,\, \xrightarrow{r\rightarrow0} \,\,  -\frac{l}{a^2}\, .
\ee
Hence, coming back to the continuation of $(\dot{\rho}, \dot{\phi})$ at the ring, it seems that geodesics have to be extended to the interior of the ring. Therefore, we have to study the geodesic equation for particles moving inside the ring. 

As we know, the Boyer-Lindquist coordinates can be expressed in terms of cylindrical coordinates as follows, 
\begin{align}\label{cylinderical}
&r=\sqrt{\frac{1}{2} (\rho^2+z^2-a^2+\sqrt{(\rho^2+z^2-a^2)^2+4a^2z^2}) }\, , \nonumber\\
&{\rm cos}\theta =\frac{z}{r}\, .
\end{align}
Inside the ring and on the equatorial plane, the coordinate $r$ is identically $0$ everywhere, while the variable ${\rm cos} (\theta)$ can be obtained by taking the limit $z\rightarrow0$ under the condition $\rho<a$, namely
\be
{\rm cos}\theta=\lim_{z\rightarrow0}\left(\frac{\partial r}{\partial z}\right)^{-1}=\frac{\sqrt{a^2-\rho^2}}{a}\, .
\ee
Therefore, since $dr=0$, $d\theta=d\rho/(a\,{\rm cos}\theta)$, and $r=0$, the metric inside the ring reads:
\be
ds^2= dt^2 - d\rho^2 - \rho^2d\phi^2\, ,
\ee
which is definitely flat. Hence, the trajectories of particles propagating inside the ring on the equatorial plane are straight lines in Cartesian coordinates.

\begin{figure}[hbt]
\centering
\begin{tikzpicture}[line width=1 pt, scale=0.75]

\draw[->,thin] (-2.5,0)--(2.6,0); \node at (2.5,0.2) {$x$}; \node at (2,2) {$r>0$};
\draw[->,thin] (0,-2.4)--(0,2.5); \node at (0.2,2.4) {$y$};
\draw (0,0) circle(1.5);
\draw [draw=red, postaction={decorate}, decoration={markings, mark=at position .55 with {\arrow[draw=red]{stealth}}}] (0,1.5)--(1.5,0);
\draw [draw=red, postaction={decorate}, decoration={markings, mark=at position .55 with {\arrow[draw=red]{stealth}}}](-2,1.5) to[out = 30, in=135] (0,1.5);
\draw [draw=red, postaction={decorate}, decoration={markings, mark=at position .55 with {\arrow[draw=red]{stealth}}}](1.5,0) to[out = -45, in=90] (2,-1.5);

\end{tikzpicture}
\caption{The trivial extension of the geodesics restricted on the equatorial plane. The circle denotes the regular ring located in $\rho=a$.}
\label{fig2}
\end{figure}
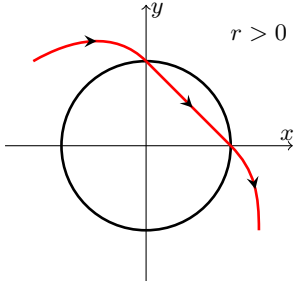

Based on the analysis above, the trajectory of the geodesics crossing the ring will be like the curve shown schematically in FIG.\ref{fig2}. Obviously, the geodesics can not be analytic curves in the two-dimensional space $\mathbb{R}^2$.
 
Therefore, if analyticity of geodesics is required, we can not extend them from the exterior to the interior of the ring, and vice versa. 

\begin{figure}[hbt]
\centering
\begin{tikzpicture}[line width=1 pt, scale=0.75]

\draw[->,thin] (-2.5,0)--(2.6,0); \node at (2.5,0.2) {$x$}; \node at (2,2) {\footnotesize $r>0$};
\draw[->,thin] (0,-2.4)--(0,2.5); \node at (0.2,2.4) {$y$}; \node at (0.7,0.6) {\footnotesize $r=0^+$};
\draw (0,0) circle(1.5);
\draw [draw=red, postaction={decorate}, decoration={markings, mark=at position .55 with {\arrow[draw=red]{stealth}}}](-2.2,0.2) to[out = 60, in=180] (135:1.5);
\draw [draw=blue, postaction={decorate}, decoration={markings, mark=at position .55 with {\arrow[draw=blue]{stealth}}}](-0.5,-0.5)--(0,-1.5);
\node at (-2,1) {\color{red}$\gamma_1$}; \node at (-0.8,-0.5) {\color{blue}$\gamma_2$};

\begin{scope}[shift={(6,0)}]
\draw[->,thin] (-2.5,0)--(2.6,0); \node at (2.5,0.2) {$x’$}; \node at (2,2) {\footnotesize $r<0$};
\draw[->,thin] (0,-2.4)--(0,2.5); \node at (0.25,2.4) {$y’$}; \node at (0.7,0.6) {\footnotesize $r=0^-$};
\draw (0,0) circle(1.5);
\draw [draw=red, postaction={decorate}, decoration={markings, mark=at position .55 with {\arrow[draw=red]{stealth}}}](135:1.5) to[out = 90, in=-135] (-0.2,2.2);
\draw [draw=blue, postaction={decorate}, decoration={markings, mark=at position .55 with {\arrow[draw=blue]{stealth}}}](0,-1.5)--(0.5,-0.5);
\node at (-0.9,2) {\color{red}$\gamma_1$}; \node at (0.8,-0.5) {\color{blue}$\gamma_2$};
\end{scope}

\end{tikzpicture}
\caption{A proposal for an analytic extension of the geodesics restricted on the equatorial plane and reaching the ring from outside ($\gamma_1$) and inside ($\gamma_2$). 
The geodesics reaching the ring from the outside will be extended to negative values of $r$, while the geodesic equations $\dot{t}$, $\dot{r}$, and $\dot{\phi}$ will be the same. The geodesics reaching the ring from inside the ring will remain inside, while the $\dot{t}$, $\dot{\phi}$ are preserved, but $\dot{\rho}$ changes the sign. Notice that here we regard $r=0^+$ and $r=0^-$ as different surfaces of the disk in $r=0$. }
\label{fig3}
\end{figure}
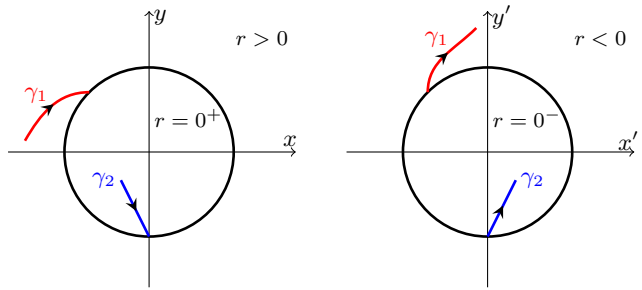

However, we do not give and we try to propose another scenario consistent with analytically. In the picture FIG.\ref{fig3}), one can imagine the ring to be the boundary between the $r>0$ and $r<0$ branches on the equatorial plane.
%
Since the geodesics in the $r>0$ and $r<0$ regions obey the same equations (\ref{eomdotr}, \ref{eomdotphi}), the extended geodesics will be analytic. 
Nevertheless, our analysis is only limited to the equatorial plane, thus, further studies are needed in order to prove the analyticity in all spacetime. Therefore, at the moment we can not secure the validity of the above extension. Furthermore, as we will see below the curvature invariants with more derivatives are singular in $r=0$, which is in evident tension with the here proposed statement of analyticity.

%
%
%
Indeed, in Ref.\cite{Modesto:2010rv, Bambi:2013ufa, Smailagic:2010nv} was already noticed that, for some RRBHs, the limit of the Kretschmann invariant at the ring depends on the direction along which we approach it. For instance, if $\tilde{m}(r)\sim r^3$ near $r=0$, we get the following two different limits of the Kretschmann scalar, \begin{align}
\lim_{\theta\rightarrow \pi/2}\left(\lim_{r\rightarrow 0} R_{\mu\nu\rho\sigma}R^{\mu\nu\rho\sigma} \right)&=0 \, ,\nonumber\\
\lim_{r\rightarrow 0}\left(\lim_{\theta\rightarrow \pi/2} R_{\mu\nu\rho\sigma}R^{\mu\nu\rho\sigma} \right) &= \frac{96M^2}{L^6} \, ,
\end{align}
which is tantamount to say the metric is not analytic at the ring. Such non-analyticity was noted in \cite{Smailagic:2010nv} and attributed
to the presence of a rotating string with Planck tension at the ring and representing the inner engine driving the rotation of all the surrounding matter.

The latter property can also be verified by calculating the higher derivatives of the metric in cylindrical coordinates. 
Indeed, assuming again $\tilde{m}\sim r^3$, one find that the second derivative of the function $\tilde{f}(\rho,z)$ (\ref{fh}), now expressed in cylindrical coordinates, diverges when approaching the ring in $\rho=a$, namely
\be
\lim_{\rho\rightarrow a^+} \left(\frac{\partial}{\partial z}\right)^2 \tilde{f}(\rho,z) \bigg |_{z=0}= \infty \, .
\ee
\be
\lim_{z\rightarrow 0} \left(\frac{\partial}{\partial \rho}\right)^2 \tilde{f}(\rho,z) \bigg |_{\rho=a}= \infty \, .
\ee

In a covariant fashion, we can infer about the presence of analyticity defects by computing higher derivative curvature invariants. For instance, for $\tilde{m}(r)\sim r^3$ near $r=0$, we have
\be\label{boxr}
\lim_{\theta\rightarrow \pi/2}\Box R \underset{r\rightarrow0}{ \sim} r^{-4}\, , 
\ee
which is divergent in $r=0$. 

At quantum level it turns out that if the classical gravitational action contains higher-derivative curvature invariants, like $\mathcal{R}\Box^n\mathcal{R}$ and $\Box^nR$, in which $\mathcal{R}$ generically denotes curvature tensors and scalars, the RRBHs' metrics might not contribute to the Lorentzian path, but filtered out by a {\em finite action principle} \cite{nature, Borissova:2020knn, Giacchini:2021pmr}.


\section{non-analytic rotating black holes}

The previous part of the paper only involves metrics that are analytic functions of the radial coordinate $r$. In this section, we briefly discuss RRBHs defined by non-analytic mass functions. 

As a typical example, we remind the spherically symmetric Culetu-Simpson-Visser metric\cite{Culetu:2013fsa, Culetu:2014lca, Simpson:2019mud, Xiang:2013sza}, whose metric is defined by the function:
\be
f(r)=1-\frac{2M}{r}e^{-k/r}\, ,
\ee
in which $k$ is a constant. The metric has a Minkowski core and the mass function is a non-analytic but $C^{\infty}$ smooth function of $r$. 
Applying the complexification (\ref{fh}), the function of the rotating Culetu-Simpson-Visser metric turns into (also see in Ref.\cite{Ghosh:2014pba})
\be
\tilde{f}(r,\theta)=1-\frac{2Mr}{\Sigma}e^{-k/r}\, .
\ee
Since the metric is not analytic at the disk in $r=0$, does not make sense to look for an analytic extension for geodesics reaching the disk. 
On the other hand, if we demand the geodesics to be $C^{\infty}$ smooth functions, the extension of the geodesics is not unique. For example, we can choose to extend the spacetime to negative values of $r$, where the spacetime is simply the Minkowski metric in $r<0$. Or, we do not extend the spacetime to $r<0$, but simply make the antipodal extension of the geodesics crossing the disk.

\section{Summary and Discussion}

We have investigated the analytic extension of the geodesics in the rotating regular black holes obtained throughout the Newman-Janis algorithm applied to spherically symmetric regular black holes.

Inspired by the maximally extended Kerr spacetime, if the metric is not symmetric under the reflection 
$r\leftrightarrow-r$, we have shown that the spacetime of RRBHs have to be analytically extended to $r<0$ too. 
 
It turns out that the geodesics passing through the interior of the ring located in $(r=0,\theta=\pi/2)$ 
are ill-defined for some RRBHs in the region $r<0$. 
For instance, the maximally extended rotating Hayward black hole, the rotating renormalization group improved black hole, and the rotating Dymnikova black hole are all singular somewhere for $r<0$. A similar issue was discovered in Ref.\cite{Zhou:2022yio} in the attempt to extend the regular spherically symmetric black holes beyond $r=0$. Other black holes likely do not suffer from this problem\cite{Bardeen, Nicolini:2005vd, Modesto:2010uh, Modesto:2010rv}. 

A much more serious problem emerges when we try to extend the geodesics reaching the ring. Indeed, since the ring singularity seems to be removed in RRBHs and the geodesics can reach the ring in a finite amount of the affine parameter, we can not avoid to investigate the subsequent evolution. 
%
For the special case of the geodesics on the equatorial plane, it may be possible to get an analytic continuation throughout an extension to negative values of $r$ at the ring. 
However, the ring turns out to be a kind of defect in which higher-derivative curvature invariants diverge. 
Therefore, all the proposed RRBHs based on analyticity turn out to be geodetically incomplete. Namely, there is no extension of the spacetime consistent with the analyticity of the geodesic equations. 
At the quantum level, it deserves to be noticed that according to the action natural selection principle \cite{nature, Borissova:2020knn} such RRBHs metrics do not contribute to the path integral whether the action contains higher-derivative curvature invariants \cite{Stelle:1976gc, Asorey:1996hz, Modesto:2011kw}. This is a generalization of the result in Ref.\cite{Giacchini:2021pmr}, where the authors focused on the spherically symmetric regular black holes and showed that all higher-derivative curvature invariants are finite if and only if the metric is symmetric under $r\leftrightarrow - r$. However, the latter regularity condition does not work for the rotating black holes defined by (\ref{rbh},\ref{fh}), namely the condition to be symmetric under $r\leftrightarrow - r$ is not sufficient to guarantee the finiteness of the curvature invariant. 





A possible way to avoid the singularity in higher derivative curvature operators can be achieved by replacing the constant angular momentum $a$ with a function of the radial coordinate, for example 
\be
a \,\, \rightarrow \,\, a' = a \, \frac{r^4}{r^4+L^4} \, .
\label{asme}
\ee
There are also some other examples to smear out the angular momentum of RRBHs\cite{Bambi:2013ufa, Casadio:2023iqt}. This kind of replacement makes the metric approach a spherically symmetric one near the ring, which now shrinks to a point. Now, since $a'(r)$ is a function of the radial coordinate that tends to zero for $r\rightarrow0$, the conclusions of Ref.\cite{Giacchini:2021pmr} applies again because the metric turns into a spherically symmetric one near $r=0$. Namely, the condition of invariance under $r \rightarrow -r$ ensures the regularity of the higher derivative curvature invariants.

Apart from the above smearing the angular momentum (\ref{asme}), one can also construct analytic and geodetically complete rotating black holes in conformal gravity (see for example \cite{Bambi:2016wdn, Modesto:2021wdd}). In such spacetimes, the particles can not reach the ring, and the nonanalyticity at the ring together with the potential singularities in $r<0$ are unattainable. It is also possible to construct a conformally rescaled rotating black hole in which the Cauchy horizon is moved to the causal infinity of the spacetime. This kind of black hole should also be able to avoid the instability issue\cite{Poisson:1990eh, Ori:1991zz} at the Cauchy horizon. By the way, as shown in Ref.\cite{Franzin:2022wai, Ghosh:2022gka}, a stable RRBH can also be obtained by applying a special mass function $m(r)$ which leads to a zero surface gravity of the Cauchy horizon making the black hole stable, and the regularity of this kind of black hole can be ensured by conformal rescaling, which pushes the disk (or the ring) to the infinity.


Finally, another proposal to avoid the singularity issue in RRBHs consists in introducing non-analytic but smooth functions of the radial coordinate $r$ \cite{Culetu:2013fsa, Simpson:2019mud}. All higher-derivative curvature invariants are finite approaching the ring, but the price to pay is that we cannot demand the analyticity of the extension anymore. In other words, the extension of geodesics in this kind of spacetimes is not unique.


\section*{ACKNOWLEDGMENTS}
This work has been supported by the Basic Research Program of the Science, Technology, and Innovation Commission of Shenzhen Municipality (grant no. JCYJ20180302174206969). 



\begin{thebibliography}{99}


\bibitem{Bardeen}
J.~M.~Bardeen, in 
``Conference Proceedings of GR5 (Tbilisi, USSR, 1968)", p. \textbf{174}.

\bibitem{Hayward:2005gi}
S.~A.~Hayward,
``Formation and evaporation of regular black holes,''
Phys. Rev. Lett. \textbf{96}, 031103 (2006).
[arXiv:gr-qc/0506126 [gr-qc]].

\bibitem{Dymnikova:1992ux}
I.~Dymnikova,
``Vacuum nonsingular black hole,''
Gen. Rel. Grav. \textbf{24}, 235-242 (1992).


\bibitem{Culetu:2013fsa}
H.~Culetu,
``On a regular modified Schwarzschild spacetime,''
arXiv:1305.5964 [gr-qc]].

\bibitem{Culetu:2014lca}
H.~Culetu,
``On a regular charged black hole with a nonlinear electric source,''
Int. J. Theor. Phys. \textbf{54}, no.8, 2855-2863 (2015)
[arXiv:1408.3334 [gr-qc]].

\bibitem{Xiang:2013sza}
L.~Xiang, Y.~Ling and Y.~G.~Shen,
``Singularities and the Finale of Black Hole Evaporation,''
Int. J. Mod. Phys. D \textbf{22}, 1342016 (2013)
[arXiv:1305.3851 [gr-qc]].



\bibitem{Simpson:2019mud} 
A.~Simpson and M.~Visser,
``Regular black holes with asymptotically Minkowski cores,''
Universe \textbf{6}, no.1, 8 (2019).

\bibitem{Bonanno:2000ep}   
A.~Bonanno and M.~Reuter,
``Renormalization group improved black hole space-times,''
Phys. Rev. D \textbf{62}, 043008 (2000); 
A.~Bonanno and M.~Reuter,
``Spacetime structure of an evaporating black hole in quantum gravity,''
Phys. Rev. D \textbf{73}, 083005 (2006).

\bibitem{Nicolini:2005vd}   
P.~Nicolini, A.~Smailagic and E.~Spallucci,
``Noncommutative geometry inspired Schwarzschild black hole,''
Phys. Lett. B \textbf{632}, 547 (2006),
gr-qc/0510112.


\bibitem{Modesto:2010uh} 
L.~Modesto, J.~W.~Moffat and P.~Nicolini,
``Black holes in an ultraviolet complete quantum gravity,''
Phys. Lett. B \textbf{695}, 397-400 (2011)
[arXiv:1010.0680 [gr-qc]].


\bibitem{Burzilla:2020utr}
N.~Burzill\`a, B.~L.~Giacchini, T.~d.~Netto and L.~Modesto,
``Higher-order regularity in local and nonlocal quantum gravity,''
Eur. Phys. J. C \textbf{81}, no.5, 462 (2021)
[arXiv:2012.11829 [gr-qc]].




\bibitem{Modesto:2008im}
L.~Modesto,
``Semiclassical loop quantum black hole,''
Int. J. Theor. Phys. \textbf{49}, 1649-1683 (2010)
[arXiv:0811.2196 [gr-qc]].

\bibitem{Modesto:2009ve}  
L.~Modesto and I.~Premont-Schwarz,
``Self-dual Black Holes in LQG: Theory and Phenomenology,''
Phys. Rev. D \textbf{80}, 064041 (2009)
[arXiv:0905.3170 [hep-th]].






\bibitem{Zhou:2022yio}
T.~Zhou and L.~Modesto,
``Geodesic incompleteness of some popular regular black holes,''
Phys. Rev. D \textbf{107}, no.4, 044016 (2023)
[arXiv:2208.02557 [gr-qc]].




\bibitem{newman}
E.~T.~Newman and A.~I.~Janis,
J. Math. Phys. \textbf{6} 915 (1965).

\bibitem{Drake:1998gf}
S.~P.~Drake and P.~Szekeres,
``Uniqueness of the Newman-Janis algorithm in generating the Kerr-Newman metric,''
Gen. Rel. Grav. \textbf{32}, 445-458 (2000)
[arXiv:gr-qc/9807001 [gr-qc]].

\bibitem{Modesto:2010rv}
L.~Modesto and P.~Nicolini,
``Charged rotating noncommutative black holes,''
Phys. Rev. D \textbf{82}, 104035 (2010)
[arXiv:1005.5605 [gr-qc]].

\bibitem{Bambi:2013ufa}
C.~Bambi and L.~Modesto,
``Rotating regular black holes,''
Phys. Lett. B \textbf{721}, 329-334 (2013)
[arXiv:1302.6075 [gr-qc]].


\bibitem{Smailagic:2010nv}
A.~Smailagic and E.~Spallucci,
``'Kerrr' black hole: the Lord of the String,''
Phys. Lett. B \textbf{688}, 82-87 (2010)
[arXiv:1003.3918 [hep-th]].



\bibitem{Casadio:2023iqt}
R.~Casadio, A.~Giusti and J.~Ovalle,
``Quantum rotating black holes,''
[arXiv:2303.02713 [gr-qc]].

\bibitem{nature}
J.~D.~Barrow and F.~J.~Tipler,
``Action principles in nature,''
Nature \textbf{331}, 31-34 (1988).


\bibitem{Borissova:2020knn}
J.~N.~Borissova and A.~Eichhorn,
``Towards black-hole singularity-resolution in the Lorentzian gravitational path integral,''
Universe \textbf{7}, no.3, 48 (2021)
[arXiv:2012.08570 [gr-qc]].


\bibitem{Giacchini:2021pmr}
B.~L.~Giacchini, T.~d.~Netto and L.~Modesto,
``Action principle selection of regular black holes,''
Phys. Rev. D \textbf{104}, no.8, 084072 (2021)
[arXiv:2105.00300 [gr-qc]].




\bibitem{Toshmatov:2014nya}
B.~Toshmatov, B.~Ahmedov, A.~Abdujabbarov and Z.~Stuchlik,
``Rotating Regular Black Hole Solution,''
Phys. Rev. D \textbf{89}, no.10, 104017 (2014)
[arXiv:1404.6443 [gr-qc]].



\bibitem{Buoninfante:2022ild}
L.~Buoninfante, B.~L.~Giacchini and T.~de Paula Netto,
``Black holes in non-local gravity,''
[arXiv:2211.03497 [gr-qc]].

\bibitem{Ghosh:2014pba}
S.~G.~Ghosh,
``A nonsingular rotating black hole,''
Eur. Phys. J. C \textbf{75}, no.11, 532 (2015)
[arXiv:1408.5668 [gr-qc]].


\bibitem{Stelle:1976gc}
K.~S.~Stelle,
``Renormalization of Higher Derivative Quantum Gravity,''
Phys. Rev. D \textbf{16}, 953-969 (1977).

\bibitem{Asorey:1996hz}
M.~Asorey, J.~L.~Lopez and I.~L.~Shapiro,
``Some remarks on high derivative quantum gravity,''
Int. J. Mod. Phys. A \textbf{12}, 5711-5734 (1997)
[arXiv:hep-th/9610006 [hep-th]].

\bibitem{Modesto:2011kw}
L.~Modesto,
``Super-renormalizable Quantum Gravity,''
Phys. Rev. D \textbf{86}, 044005 (2012)
[arXiv:1107.2403 [hep-th]].

\bibitem{Bambi:2016wdn}
C.~Bambi, L.~Modesto and L.~Rachwa\l{},
``Spacetime completeness of non-singular black holes in conformal gravity,''
JCAP \textbf{05}, 003 (2017)
[arXiv:1611.00865 [gr-qc]].

\bibitem{Modesto:2021wdd}
L.~Modesto, A.~Akil and C.~Bambi,
``Conformalons and Trans-Planckian problem,''
[arXiv:2106.03914 [gr-qc]].

\bibitem{Poisson:1990eh}
E.~Poisson and W.~Israel,
``Internal structure of black holes,''
Phys. Rev. D \textbf{41}, 1796-1809 (1990); 
\bibitem{Ori:1991zz}
A.~Ori,
``Inner structure of a charged black hole: An exact mass-inflation solution,''
Phys. Rev. Lett. \textbf{67}, 789-792 (1991).


\bibitem{Franzin:2022wai}
E.~Franzin, S.~Liberati, J.~Mazza and V.~Vellucci,
``Stable rotating regular black holes,''
Phys. Rev. D \textbf{106}, no.10, 104060 (2022)
[arXiv:2207.08864 [gr-qc]].


\bibitem{Ghosh:2022gka}
R.~Ghosh, M.~Rahman and A.~K.~Mishra,
``Regularized stable Kerr black hole: cosmic censorships, shadow and quasi-normal modes,''
Eur. Phys. J. C \textbf{83}, no.1, 91 (2023)
[arXiv:2209.12291 [gr-qc]].


\end{thebibliography}
\end{document}